\documentclass[11pt]{amsart}
\usepackage{fullpage,amsmath,amsfonts,amssymb,amsthm,epsfig,color, graphicx, natbib, appendix,comment,subfigure, graphicx, amssymb,xcolor}

\usepackage[colorlinks=true,
    linkcolor=blue,
    filecolor=magenta,      
    urlcolor=blue]{hyperref}
\usepackage{tikz}
\usetikzlibrary{shapes,arrows}
\usetikzlibrary{intersections}
\usepackage[capitalise,noabbrev,nameinlink]{cleveref}

\usepackage[T1]{fontenc}
\usepackage{fourier}

\usepackage[margin=1in]{geometry}

\newtheorem*{definition}{Definition}

\definecolor{darkblue}{rgb}{0.0, 0.0, 0.55}
\hypersetup{colorlinks,linkcolor={darkblue},citecolor={darkblue},urlcolor={darkblue}} 

\title{On the difficulty of characterizing  \\ network formation with endogenous behavior}

\author{Benjamin Golub \and Yu-Chi Hsieh \and Evan Sadler}

\thanks{Golub: ben.golub@gmail.com, Northwestern University. Hsieh: yuchi.hsieh@kellogg.northwestern.edu, Northwestern University. Sadler: es3668@columbia.edu, Columbia University. We are grateful to Ugo Bolletta for helpful conversations; all errors are our own.}

\begin{document}

\begin{abstract}
\citet{bolletta2021model} studies a model in which a network is strategically formed and then agents play a linear best-response investment game in it. The model is motivated by an application in which people choose both their study partners and their levels of educational effort. Agents have different one-dimensional types---private returns to effort. A main result claims that (pairwise Nash) stable networks have a \emph{locally complete} structure consisting of possibly overlapping cliques: if two agents are linked, they are part of a clique composed of all agents with types between theirs. A counterexample shows that the claimed characterization is incorrect. We specify where the analysis errs and discuss implications for network formation models.
\end{abstract}

\date{\today}
\maketitle

\cite{bolletta2021model} studies the joint strategic choice of network links and productive effort, motivated by applications in the economics of education. The theoretical findings are used in interpreting  empirical studies on peer effects, such as \citet{carrell2013natural}. This fits into a literature on network games\footnote{For surveys, see \citet*{kranton_survey} and \citet*{jackson_zenou}.} and  games on endogenous networks---see, e.g.,  \citet{Konigetal2014}, \cite{Hiller2017}, and \citet{Badev2021}.

The paper examines a stylized network formation process. At each time, starting from a status quo network, two (random) agents $i$ and $j$ can form a new link between them by mutual consent; alternatively, either can unilaterally sever any of their own existing links. After each such stage, agents play the static Nash equilibrium of a coordination game in the current network: each wants to match an average of a private ideal effort (its ``type'') and average behavior in its neighborhood. Higher types are inclined to make higher efforts in the effort game, holding spillovers fixed.   In assessing the consequences of their decisions in the formation stage, players anticipate equilibrium behavior in the action-choice game on the resulting network.  The process is said to reach an equilibrium if no pair of agents wish to change their linking decisions at the formation stage. In the leading application, links are study-partner relationships and the action is educational efforts.

A basic theoretical question is: which networks can be equilibrium outcomes?  A main claim of \cite{bolletta2021model}, stated in its Proposition 1, is that equilibrium networks have a specific structure, called \emph{locally complete}. This means that, at any equilibrium, if two agents are connected, they and all agents with types between theirs form a clique---a completely connected subnetwork.

The purpose of this note is to point out that this proposition is false. We do this by providing an explicit counterexample: an equilibrium outcome that is not locally complete. We also show that this outcome can be reached by the dynamic process studied in \citet{bolletta2021model} starting from the empty network. It is thus a natural outcome reachable from a ``neutral'' status quo. Our note also examines arguments that underlie the proposition in order to highlight where the analysis errs.

We give a brief roadmap. In \cref{sec:model}, we establish key notation and review Bolletta's model as well as some definitions relevant  both to Bolletta's and our analyses. In \cref{sec:counterexample}, we present our counterexample to Proposition 1 in the paper; we also discuss the key errors in the proof. \cref{sec:dynamic} shows that the counterexample network is a natural outcome of the dynamic process, not just an equilibrium of it. \cref{sec:conclusion} concludes.

\section{The model and a main result of \cite{bolletta2021model}} \label{sec:model}
In this section, we review \citet{bolletta2021model}'s model setup and state the result that we focus on. We mostly work within the same notation, but in some cases we make slight modifications.

\subsection{Model} There are $N$ agents. All networks discussed are undirected, unweighted graphs without self-edges on the nodes $[N]:=\{1,2,\ldots,N\}$. Agent $i$ has a type $\theta_i$, a neighborhood $g_i\subseteq [N]$, and an action $y_i$. Given a network $g$, the degree of player $i$ is $d_i = |g_i|$, the number of links $i$ has. The vector of efforts for all players is denoted by $y = (y_1, \ldots, y_N)$. The payoff of player $i$ is given by
$$
U_i = u\left(\mathbf{y}, \theta_i, g\right)=(1-\alpha) \theta_i y_i  - \frac{y_i^2}{2} +
\sum_{j \in g_i} \left( \delta + \frac{\alpha y_i y_j}{d_i} \right)
$$
if player $i$ has a non-empty neighborhood, and $U_i = \theta_i y_i - \frac{y_i^2}{2}$ otherwise. The value $\delta$ represents an exogenous positive benefit that a player receives from each link. The parameter $\alpha \in [0,1)$ captures the weight that players put on matching their own type versus the average of their neighbors' actions. We assume that $\theta_i$ can take on any positive real value. By standard results on network games, if $\alpha \in [0,1)$, then fixing a network $g$, the game of effort choice has a unique equilibrium \citep[Lemma 1]{bolletta2021model}.

The paper proceeds to introduce a dynamic process of network evolution and an equilibrium concept, which defines points at which the process can come to rest. The main claims of Bolletta's paper apply to the equilibrium outcome,  and so we can present our counterexample using only the definitions above. Later, in  \cref{sec:dynamic}, we discuss the dynamic process leading to equilibrium outcomes and discuss our counterexample in light of it.

\subsection{Equilibria} \label{sec:equilibria} Given a specification of payoffs in each possible network, an equilibrium network is defined to be one in which no pair of agents mutually wishes to create an additional link, and no agent wants to unilaterally sever any subset of her existing links. This equilibrium concept corresponds to the standard notion of pairwise Nash stability. Crucially, the payoffs of any network $g'$ (contemplated by agents at the status quo and in their deviations) are determined by equilibrium effort choices $y^*(g')$ given the network $g'$. We denote these payoffs by $U_i(g')$.
\begin{definition}[Pairwise Nash stability (\cite{jackson2010social})] A network $g$ is pairwise Nash stable if
\begin{enumerate}
    \item[1.] For all $i j \notin g$, if $U_i(g+i j)>U_i(g)$ then $U_j(g+i j)<U_j(g)$.
    \item[2.] For all $i\in N$ and any set $L$ of edges involving $i$, we have $U_i(g)\geq U_i(g-L)$.\footnote{Equivalently, the network $g$ is a Nash equilibrium of the Myerson link announcement game, in which all agents simultaneously announce a desired set of links, and the link $ij$ forms if both $i$ and $j$ announced it.}
\end{enumerate}
\end{definition}

The specification of what payoffs agents anticipate when contemplating deviations from $g$ is an important modeling choice. We have made the same one as \citet[Definition 1]{bolletta2021model} (and we have confirmed in correspondence with the author that our interpretation is the intended one). An alternative assumption would be that effort choices $y$ are fixed at the current levels (equilibrium efforts given $g$).\footnote{There are also other possibilities, involving more farsighted anticipation of where further deviations may lead, but \cite{bolletta2021model} does not consider this possibility, and we will not, either.}    Some further discussion of this can be found in \cref{sec:other_issues}, where we show that some of the key arguments of \citet{bolletta2021model} do not hold under either modeling choice.

Next, we define a locally complete network, which is a notion central to Proposition 1 of  \citet{bolletta2021model}.  This definition is adapted to the present setting from \cite{dutta2013networks}.

\begin{definition}[Locally complete network] 
Let $i \leftrightarrow j:=\{k \in [N] \mid \theta_i \leq \theta_k \leq \theta_j\}$ be the set of all agents $k$ such that $\theta_k$ is in the interval $[\theta_i,\theta_j]$ (including $i$ and $j$ themselves). A network $g$ is called \emph{locally complete} if, whenever $\theta_i \leq \theta_j$ and $ij\in g$, the graph $g$ contains the complete graph on $i \leftrightarrow j$.
\end{definition}

\subsection{Result} The main result of \cite{bolletta2021model} asserts a characterization of pairwise Nash stable outcomes as defined above. The result is stated as Proposition 1:
\begin{quote}
\emph{Every stable network is a locally complete network.}
\end{quote}

\section{A Counterexample to Bolletta's Proposition 1} \label{sec:counterexample}

\begin{figure}[htbp]
\centering
\input{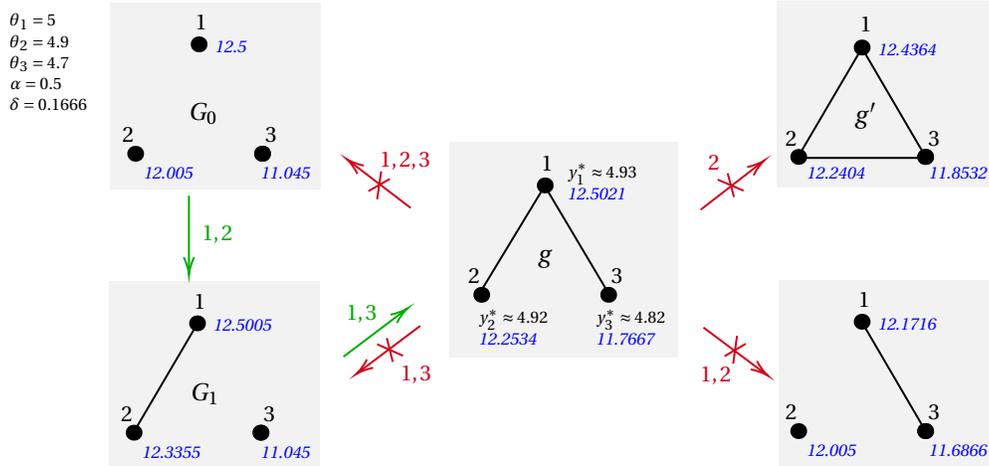}
\caption{\footnotesize A counterexample to Bolletta's Proposition 1. The payoffs of all agents (under equilibrium actions) are shown in blue, italicized text. Crossed-out, red arrows indicate moves that at least one pivotal agent (indicated in the arrow label) does not benefit from. These are used to check stability. Green arrows are moves (link creations) from which the involved agents both weakly benefit. They are important to the reachability of $g$ and are discussed in \cref{sec:reachability}.
}
\label{fig:counterexample}
\end{figure}

In this section, we provide a counterexample to Proposition 1 in \cite{bolletta2021model}. We consider three-player networks with the parameters, agents' equilibrium actions ($y^*_i$, in black) and utilities (in blue, italicized) shown in \cref{fig:counterexample}. 

We will show network $g = \{ \{1, 2\}, \{1, 3\} \}$ is pairwise Nash stable. Note this network is not locally complete:  agent 2 with the middle type is not linked to agent 3.\footnote{ More precisely, note agent 1 and agent 3 are linked and $\theta_2 \in (\theta_1, \theta_3)$, while agent 2 and agent 3 are not linked.}

To show pairwise stability, we consider all networks in $\mathcal{G}(\{1,2,3\})$ that can be reached by a deviation considered in the definition of pairwise Nash stability.
As shown in \cref{fig:counterexample}, no deviation that can be effected unilaterally or by a pair of agents is profitable. 
Thus, $g$ is a pairwise Nash stable network but, as we have said, not a locally complete one.  This contradicts Bolletta's Proposition 1. %

It is worth commenting on the proof of Proposition 1 and where it goes wrong when applied to the counterexample. A key step in the proof asserts (without argument) that if there are three agents with distinct types ordered  $\theta_3 < \theta_2 < \theta_1$, and agent $1$ is connected to both agents $2$ and $3$, then agent $2$ must also want to link with agent $3$.%
It may seem intuitive that if even the high type wants to link with the low type, then the medium type would also want to. Our counterexample shows that this need not be so: in the move $g\to g'$ in \cref{fig:counterexample}, agent $2$ is connected with agent $1$ but prefers not to link with agent $3$. Intuitively, the flaw in the claim is that it overlooks the high- and medium-type agents' different neighborhoods and the impact of adding link $\{1,3\}$ on actions.

\subsection{A key monotonicity lemma and a counterexample} \label{sec:other_issues}

Beyond the error we have just discussed, a key building block in Bolletta's proof of Proposition 1 is a lemma---potentially of independent interest---asserting that agents' preferences over partners are monotonic in their peers' types. More precisely, the paper's Lemma 2 states:
\begin{quote}\emph{Consider the well-ordered set of agents $\Theta:=\left\{\theta_1, \ldots\right.$, $\left.\theta_i, \ldots, \theta_n \mid \theta_1<\cdots<\theta_i<\cdots<\theta_n\right\}$. For all agents $i, U_i(y, g+i k)>$ $U_i(y, g+i j)$ for all $\theta_k>\theta_j$. Moreover, $y_i^*(g+i k)>y_i^*(g+i j)$, for each $\alpha \in[0,1)$.}\end{quote}
\noindent  In the proof, a violation of local completeness branches into two cases, with the erroneous analysis we have discussed above dealing with one case, and Lemma 2 dealing with the other case.  

It is intuitive that this lemma should be useful in deducing a lot of structure in equilibrium networks: each agent wants to link with all those above a certain exogenous type, which helps in ruling out ``gaps'' in neighborhoods. However, in this section, we show that this lemma is also false. As we will discuss more later, this suggests that imposing the right amount of order on the formation process is trickier than it may appear.

\begin{figure}[htbp] \centering
\tikzset{every picture/.style={line width=0.75pt}} %

\begin{tikzpicture}[x=0.75pt,y=0.75pt,yscale=-1,xscale=1]

\draw  [color={rgb, 255:red, 0; green, 0; blue, 0 }  ][line width=6] [line join = round][line cap = round] (250.56,142) .. controls (250.56,142) and (250.56,142) .. (250.56,142) ;
\draw  [color={rgb, 255:red, 0; green, 0; blue, 0 }  ][line width=6] [line join = round][line cap = round] (389.56,192) .. controls (389.56,192) and (389.56,192) .. (389.56,192) ;
\draw  [color={rgb, 255:red, 0; green, 0; blue, 0 }  ][line width=6] [line join = round][line cap = round] (268.56,191) .. controls (268.56,191) and (268.56,191) .. (268.56,191) ;
\draw  [color={rgb, 255:red, 0; green, 0; blue, 0 }  ][line width=6] [line join = round][line cap = round] (410.56,142) .. controls (410.56,142) and (410.56,142) .. (410.56,142) ;
\draw    (251,142) -- (268.33,191.33) ;
\draw    (411,143) -- (389.33,192.33) ;

\draw (255.95,124.99) node [anchor=north west][inner sep=0.75pt]  [font=\footnotesize]  {$1$};
\draw (375.95,173.99) node [anchor=north west][inner sep=0.75pt]  [font=\footnotesize]  {$3$};
\draw (273.95,172.99) node [anchor=north west][inner sep=0.75pt]  [font=\footnotesize]  {$2$};
\draw (393.95,124.99) node [anchor=north west][inner sep=0.75pt]  [font=\footnotesize]  {$4$};
\draw (208,132) node [anchor=north west][inner sep=0.75pt]  [font=\fontsize{0.59em}{0.71em}\selectfont]  {$ \begin{array}{l}
\theta \mathnormal{_{1}} =20\\
y\mathnormal{_{1}^{*}} =16
\end{array}$};
\draw (226,181) node [anchor=north west][inner sep=0.75pt]  [font=\fontsize{0.59em}{0.71em}\selectfont]  {$ \begin{array}{l}
\theta \mathnormal{_{2}} =10\\
y\mathnormal{_{2}^{*}} =14
\end{array}$};
\draw (411.83,131) node [anchor=north west][inner sep=0.75pt]  [font=\fontsize{0.59em}{0.71em}\selectfont]  {$ \begin{array}{l}
\theta \mathnormal{_{4}} =13\\
y\mathnormal{_{4}^{*}} =12.2
\end{array}$};
\draw (392,181) node [anchor=north west][inner sep=0.75pt]  [font=\fontsize{0.59em}{0.71em}\selectfont]  {$ \begin{array}{l}
\theta \mathnormal{_{3}} =11\\
y\mathnormal{_{3}^{*}} =11.8
\end{array}$};

\end{tikzpicture}
\caption{\footnotesize The original equilibrium (with $\alpha=\frac{2}{3}$ and $\delta = 75$)}
\label{fig:originalequi}
\end{figure}
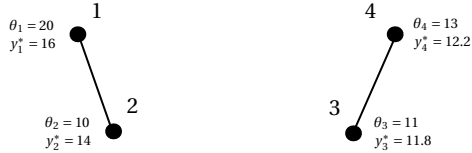

To give the counterexample, we examine an equilibrium (called the original equilibrium) of a four-player network with the parameters and equilibrium actions shown in \cref{fig:originalequi}.

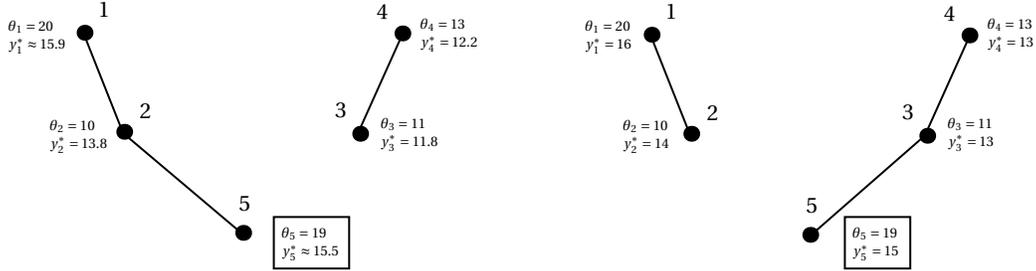
\begin{figure}[htbp] \centering
\tikzset{every picture/.style={line width=0.75pt}} %

\begin{tikzpicture}[x=0.75pt,y=0.75pt,yscale=-1,xscale=1]

\draw  [color={rgb, 255:red, 0; green, 0; blue, 0 }  ][line width=6] [line join = round][line cap = round] (100.56,120) .. controls (100.56,120) and (100.56,120) .. (100.56,120) ;
\draw  [color={rgb, 255:red, 0; green, 0; blue, 0 }  ][line width=6] [line join = round][line cap = round] (239.56,171) .. controls (239.56,171) and (239.56,171) .. (239.56,171) ;
\draw  [color={rgb, 255:red, 0; green, 0; blue, 0 }  ][line width=6] [line join = round][line cap = round] (180.56,221) .. controls (180.56,221) and (180.56,221) .. (180.56,221) ;
\draw  [color={rgb, 255:red, 0; green, 0; blue, 0 }  ][line width=6] [line join = round][line cap = round] (120.56,170) .. controls (120.56,170) and (120.56,170) .. (120.56,170) ;
\draw  [color={rgb, 255:red, 0; green, 0; blue, 0 }  ][line width=6] [line join = round][line cap = round] (260.86,120.34) .. controls (260.86,120.34) and (260.86,120.34) .. (260.86,120.34) ;
\draw    (100.33,121) -- (120,170.67) ;
\draw    (260.62,121.33) -- (238.35,170.92) ;
\draw   (196,213) -- (234,213) -- (234,239) -- (196,239) -- cycle ;
\draw    (180.33,221.67) -- (120,170.67) ;
\draw  [color={rgb, 255:red, 0; green, 0; blue, 0 }  ][line width=6] [line join = round][line cap = round] (386.56,121) .. controls (386.56,121) and (386.56,121) .. (386.56,121) ;
\draw  [color={rgb, 255:red, 0; green, 0; blue, 0 }  ][line width=6] [line join = round][line cap = round] (525.56,172) .. controls (525.56,172) and (525.56,172) .. (525.56,172) ;
\draw  [color={rgb, 255:red, 0; green, 0; blue, 0 }  ][line width=6] [line join = round][line cap = round] (466.56,222) .. controls (466.56,222) and (466.56,222) .. (466.56,222) ;
\draw  [color={rgb, 255:red, 0; green, 0; blue, 0 }  ][line width=6] [line join = round][line cap = round] (406.56,171) .. controls (406.56,171) and (406.56,171) .. (406.56,171) ;
\draw  [color={rgb, 255:red, 0; green, 0; blue, 0 }  ][line width=6] [line join = round][line cap = round] (546.86,121.34) .. controls (546.86,121.34) and (546.86,121.34) .. (546.86,121.34) ;
\draw    (386.33,122) -- (406,171.67) ;
\draw    (546.62,122.33) -- (524.35,171.92) ;
\draw   (484,213) -- (517,213) -- (517,239) -- (484,239) -- cycle ;
\draw    (466.33,222.67) -- (524.35,171.92) ;

\draw (479.83,215.43) node [anchor=north west][inner sep=0.75pt]  [font=\fontsize{0.59em}{0.71em}\selectfont]  {$ \begin{array}{l}
\theta \mathnormal{_{5}} =19\\
y\mathnormal{_{5}^{*}} =15
\end{array}$};
\draw (345,111) node [anchor=north west][inner sep=0.75pt]  [font=\fontsize{0.59em}{0.71em}\selectfont]  {$ \begin{array}{l}
\theta \mathnormal{_{1}} =20\\
y\mathnormal{_{1}^{*}} =16
\end{array}$};
\draw (364,161) node [anchor=north west][inner sep=0.75pt]  [font=\fontsize{0.59em}{0.71em}\selectfont]  {$ \begin{array}{l}
\theta \mathnormal{_{2}} =10\\
y\mathnormal{_{2}^{*}} =14
\end{array}$};
\draw (547.83,110) node [anchor=north west][inner sep=0.75pt]  [font=\fontsize{0.59em}{0.71em}\selectfont]  {$ \begin{array}{l}
\theta \mathnormal{_{4}} =13\\
y\mathnormal{_{4}^{*}} =13
\end{array}$};
\draw (527.83,160) node [anchor=north west][inner sep=0.75pt]  [font=\fontsize{0.59em}{0.71em}\selectfont]  {$ \begin{array}{l}
\theta \mathnormal{_{3}} =11\\
y\mathnormal{_{3}^{*}} =13
\end{array}$};
\draw (105.95,102.99) node [anchor=north west][inner sep=0.75pt]  [font=\footnotesize]  {$1$};
\draw (225.62,153.67) node [anchor=north west][inner sep=0.75pt]  [font=\footnotesize,rotate=-1.08]  {$3$};
\draw (176.95,200.99) node [anchor=north west][inner sep=0.75pt]  [font=\footnotesize]  {$5$};
\draw (126.95,153.99) node [anchor=north west][inner sep=0.75pt]  [font=\footnotesize]  {$2$};
\draw (246.56,104.05) node [anchor=north west][inner sep=0.75pt]  [font=\footnotesize,rotate=-1.08]  {$4$};
\draw (54,111) node [anchor=north west][inner sep=0.75pt]  [font=\fontsize{0.59em}{0.71em}\selectfont]  {$ \begin{array}{l}
\theta \mathnormal{_{1}} =20\\
y\mathnormal{_{1}^{*}} \approx 15.9
\end{array}$};
\draw (75,161) node [anchor=north west][inner sep=0.75pt]  [font=\fontsize{0.59em}{0.71em}\selectfont]  {$ \begin{array}{l}
\theta \mathnormal{_{2}} =10\\
y\mathnormal{_{2}^{*}} =13.8
\end{array}$};
\draw (261.83,110) node [anchor=north west][inner sep=0.75pt]  [font=\fontsize{0.59em}{0.71em}\selectfont]  {$ \begin{array}{l}
\theta \mathnormal{_{4}} =13\\
y\mathnormal{_{4}^{*}} =12.2
\end{array}$};
\draw (191.83,215.43) node [anchor=north west][inner sep=0.75pt]  [font=\fontsize{0.59em}{0.71em}\selectfont]  {$ \begin{array}{l}
\theta \mathnormal{_{5}} =19\\
y\mathnormal{_{5}^{*}} \approx 15.5
\end{array}$};
\draw (241.83,160) node [anchor=north west][inner sep=0.75pt]  [font=\fontsize{0.59em}{0.71em}\selectfont]  {$ \begin{array}{l}
\theta \mathnormal{_{3}} =11\\
y\mathnormal{_{3}^{*}} =11.8
\end{array}$};
\draw (391.95,103.99) node [anchor=north west][inner sep=0.75pt]  [font=\footnotesize]  {$1$};
\draw (511.62,154.67) node [anchor=north west][inner sep=0.75pt]  [font=\footnotesize,rotate=-1.08]  {$3$};
\draw (462.95,201.99) node [anchor=north west][inner sep=0.75pt]  [font=\footnotesize]  {$5$};
\draw (412.95,154.99) node [anchor=north west][inner sep=0.75pt]  [font=\footnotesize]  {$2$};
\draw (532.56,105.05) node [anchor=north west][inner sep=0.75pt]  [font=\footnotesize,rotate=-1.08]  {$4$};

\end{tikzpicture}
\caption{\footnotesize A counterexample to Bolletta's Lemma 2}
\label{fig:counterexamplelemma2}
\end{figure}

Consider agent 5, an outside agent with no pre-existing links and a type of $\theta_5=19$. Using the notation from Lemma 2, let $j=2$ and $k=3$, and compare the two equilibria depicted in \cref{fig:counterexamplelemma2}. Agent 5 would have a higher utility if it linked only with agent 2 ($y^\ast_5 \approx 15.5$) rather than if it linked only with agent 3 ($y^\ast_5 = 15$).  This contradicts the statement of the lemma.

We believe that the issue in the proof of Lemma 2 lies in the final three lines (p. 9).\footnote{Note $N_i$ is used in Bolletta's proof for $g_i$; we have replaced $N_i$ by $g_i$ in the quotation to avoid confusion.}
\begin{quote}
    \emph{Now call $\overline{y}_i'$ the term such that $k \in g_i$ and $j \notin g_i$ and $\overline{y}_i$ the term such that $j \in g_i$ and $k \notin g_i$. The simple observation that $\overline{y}_i'>\overline{y}_i$ completes the proof.}
\end{quote}

In this proof, $\overline{y}_i'$ and $\overline{y}_i$ refer to the average equilibrium actions in $i$'s neighborhood in the new networks; in more detail, these averages are equal to $\frac{1}{d_i} \sum_{j \in g_i} y_j$, for different specifications of the neighborhood $g_i$. According to this interpretation, the claim that ``$\theta_k>\theta_j$ implies $\overline{y}_i'>\overline{y}_i$ and thus $y^\ast_i(g+ik) > y^\ast_i(g+ij)$'' is shown to be false by our counterexample.

\subsubsection{An alternative specification} The model we have worked with throughout defines $U_i$ by positing that agents correctly anticipate equilibrium actions in the new network following any deviation. We have shown the lemma is false under this specification of $U_i$. But another specification that seems reasonable is one where agents make linking decisions \emph{myopically}, assuming others' actions remain the same as before the link change. (This interpretation is suggested by the quoted lines of proof above, which do not seem to consider changes in equilibrium actions after link changes.) Our counterexample disproves the lemma under this interpretation, as well. Consider the original equilibrium in \cref{fig:originalequi}: Agent 5 would prefer to link with agent 2 ($y^\ast_2=14$) rather than with agent 3 ($y^\ast_3=11$), even though $\theta_3 > \theta_2$. Thus, linking incentives are not monotonic in opponent types.

\section{A dynamic process} \label{sec:dynamic}

The main claims of \citet{bolletta2021model} can be considered solely in the context of stable outcomes, as we have done so far. However, the paper also spends considerable time on a dynamic process where a network is formed by successive rounds of (pairwise consensual) link formation and (unilateral) link deletion. Natural outcomes of such a process are ones that are not only stable once formed, but reachable through the dynamic process. This section formalizes a notion of reachability and shows that our counterexample network satisfies it. This strengthens our main result.

\subsection{A network formation process} \label{sec:definitions}

\cite{bolletta2021model} adopts a process from \cite{watts2001dynamic}, as described in Definition 2: 
\begin{quote}
\emph{Players meet over time $T = 1,2,\ldots,t,\ldots$.  At each period, a pair $ij$ is selected to decide whether to form a link or sever one that already exists. Players selected this way can simultaneously sever any existing links with all $k \in g_i$ and $h \in g_j$.}
\end{quote} 
The timing of each stage of the process (as presented in the paper) is:
\begin{quote} 
{\begin{enumerate}
\item[1.]  \emph{Agents form the network,} \item[2.] \emph{Outcomes are determined as the solution of a system of best responses.}
\end{enumerate} }
\end{quote}

We remark that there is some ambiguity in the definition of the formation process because it does not specify \emph{which} network agents think will be in effect for the purpose of determining effort choices. Will it be the network formed immediately after their deviation, or do they anticipate further deviations by others? We now fully specify the network formation process, choosing the same specification of $U_i$ as in the stability analysis. This resolution is based on our correspondence with the author of \cite{bolletta2021model}.

\medskip

\noindent \textbf{Network formation process.}
Time proceeds in discrete instants indexed by the nonnegative integers, and $(G_t)_{t=1}^{T'}$ is a sequence of $N$-player networks indexed by time $t$. We let uppercase $G$ denote networks appearing in the formation process, whereas lowercase $g$ stands for arbitrary networks. We set the initial network $G_0$ to be the empty network. At time $t$, starting from network $G_t$, the following steps occur.
\begin{enumerate}
    \item[1.] A pair of players $i$ and $j$ is (randomly) selected to act. If either player can strictly benefit by unilaterally severing some subset of her links, she does so.\footnote{In contemplating deletions, the player assumes the network is otherwise held fixed. If there are multiple subsets that a player can profitably sever, one can be chosen arbitrarily.}  In this case, the round ends. Otherwise, $i$ and $j$  form a link if doing so weakly benefits both and strictly benefits at least one. 
    \item[2.] A Nash equilibrium effort profile in the network $G_{t+1}$ is played and players receive payoffs $U_i(G_{t+1})$; these are the payoffs used in determining which deviations are profitable in (1).
\end{enumerate}

We say the sequence of networks $\{G_t\}$ \emph{reaches} an equilibrium $g$ at some finite time $T$ if $G_t = g$ for all $t\geq T$. We call such equilibrium networks \emph{reachable}.

\subsection{The counterexample network is reachable} \label{sec:reachability} 
In order to prove that the counterexample network $g$ is reachable, we reason as follows. Starting from the empty network, both agent 1 and agent 2 are better off by forming the link $\{1, 2\}$. Then, similarly, both agent 1 and agent 3 are better off by forming the link $\{1, 3\}$. Therefore, the network $g = \{\{1, 2\},\{1, 3\}\}$ is reachable through the network formation process $G_0 = \{\}\rightarrow G_1=\{\{1, 2\}\}\rightarrow G_2 = g$, illustrated by the two green arrows in \cref{fig:counterexample}. %

\section{Concluding discussion} \label{sec:conclusion}

\citet{bolletta2021model} examines a model of endogenous network formation combined with subsequent effort choice, and the paper's Proposition 1 claims equilibria necessarily involve locally complete networks. In this note, we have presented a counterexample to that result and identified an error in a crucial lemma used in the proof of the proposition.

\begin{sloppypar}
More broadly, our analysis highlights that in a setting with endogenous network formation and endogenous effort choice, clean characterizations of equilibrium outcomes are difficult to obtain. \citet{sadler2021games} makes progress on this problem, identifying  conditions that are sufficient to ensure locally complete networks. In fact, some of the key arguments there work in a similar way to the above-discussed proof of Bolletta's Lemma 2, but the analysis has to be different in several ways to make these arguments valid. For example, the  ordering in the definition of locally complete networks is based on agents' equilibrium \emph{actions}, not their types. 
\end{sloppypar}

Equally importantly, the qualitative properties of incentives in \citet{bolletta2021model} are very different from those used by Sadler and Golub to deduce locally complete structure. Under the separable spillovers studied in the latter paper, externalities from a neighbor's effort do not depend on the agent's degree. In Bolletta's coordination game, spillovers are averaged and do depend on one's degree.\footnote{The distinction between total and average spillovers shows up in other related problems as well.  For instance, \citet{Galeottietal2010} study network games of strategic complements and substitutes, and they obtain much stronger comparative statics results in examples with additive spillovers. For more on distinctions between ``local aggregate'' and ``local average'' network games---see \citet{liu2014endogenous} and \citet{boucher2022toward}.} That degree-dependence makes it hard to deduce locally complete structure.  Even if we make the definition of local completeness based on equilibrium actions, rather than types (sidestepping the lack of a monotonic relationship between actions and types that we have discussed in connection with Lemma 2), \cref{fig:counterexample} nevertheless shows that in Bolletta's model, stable networks are not locally complete. That is, two linked agents need not be part of a clique consisting of all the agents taking actions between theirs.\footnote{Agent $1$ and $3$ are linked, and agent $2$'s action is between theirs, but the three do not form a clique.}

In short, when links and behavior are both endogenous, the relationship between payoff structure and network structure is subtle. Local-average settings present considerable complications, while progress can be made under different assumptions. Though some of the results in \cite{bolletta2021model} are incorrect, in our view his paper highlights both the importance and the difficulty of the kind of problem it studies.

\bibliographystyle{ecta}
{\footnotesize \bibliography{refs}}

\end{document}